\documentstyle[aps,12pt]{revtex}
\begin{document}
\title{Toroidal dipole resonances in the relativistic
random phase approximation}
\author{D. Vretenar$^{1,2}$, N. Paar$^{2}$, P. Ring$^{2}$,
and T. Nik\v si\' c$^{1,2}$}
\address{
1. Physics Department, Faculty of Science, 
University of Zagreb, 10000 Zagreb, Croatia\\
2. Physics Department, TU Munich, D-85748 Garching,
Germany \\
}
\date{\today}
\maketitle
\bigskip
\bigskip
\begin{abstract}
The isoscalar toroidal dipole strength distributions
in spherical nuclei are calculated in the framework of a fully
consistent relativistic random
phase approximation. 
It is suggested that the recently observed
"low-lying component of the isoscalar dipole mode"
might in fact correspond to the toroidal giant dipole
resonance. Although predicted by several theoretical models,
the existence of toroidal resonances has not yet been confirmed
in experiment. The strong mixing between the toroidal 
resonance and the dipole compression mode might help to explain the 
large discrepancy between theory and experiment on the 
position of isoscalar giant dipole resonances.
\end{abstract}


%
In addition to data on isoscalar giant monopole resonances 
(IS GMR) in spherical nuclei, it is expected that information
on nuclear matter incompressibility should also be obtained
in studies of the isoscalar dipole mode. The isoscalar 
giant dipole resonance (IS GDR) is a second order
effect, built on $3\hbar \omega$, or higher configurations.
It corresponds to a compression wave traveling back
and forth through the nucleus along a definite direction. 
Recent data on IS GDR obtained by using inelastic 
scattering of $\alpha$ particles on $^{208}$Pb~\cite{Dav.97}, and
on $^{90}$Zr, $^{116}$Sn, $^{144}$Sm, and $^{208}$Pb~\cite{Cla.99},
have been analyzed in the non-relativistic Hartree-Fock plus RPA
framework~\cite{Col.00}, and with relativistic mean-field plus RPA 
(RRPA) calculations~\cite{VWR.00}.
Both analyses have shown that: (a) there is a strong disagreement
between theory and the reported experimental data 
on the position of the IS GDR centroid energies,
and (b) calculations predict the splitting of the IS GDR strength 
distribution into two broad structures, one in 
the high-energy region above 20 MeV, and one in the 
low-energy window between 8 MeV and 14 MeV. Effective interactions,
both non-relativistic and relativistic, which 
reproduce the experimental excitation energies of the IS GMR,
predict centroid energies of the IS GDR that are $4-5$ MeV higher 
than those extracted from small angle $\alpha$-scattering spectra.
This disagreement between theory and experiment is
an order of magnitude larger than for other giant resonances.
Another puzzling result is the theoretical prediction of 
a substantial amount of isoscalar dipole
strength in the $8 - 14$ MeV region. In Ref.~\cite{VWR.00} we have 
shown that the RRPA peaks in this region do not correspond
to a compression mode, but rather to a kind of toroidal motion
with dynamics determined by surface effects. 
In a very recent article~\cite{CLY.01}, Clark {\it et al.} 
reported new experimental data on the isoscalar dipole 
strength functions in $^{90}$Zr, $^{116}$Sn, and $^{208}$Pb,
measured with inelastic scattering of $\alpha$ particles at 
small angles. They found that the isoscalar E1 strength 
distribution in each nucleus consists of a broad component 
at $E_x \approx 114/A^{1/3}$ MeV containing approximately 
100\% of the E1 EWSR, and a narrower one at $E_x \approx 72/A^{1/3}$
MeV containing $15 - 28$ \% of the total isoscalar E1 strength.
The higher component is identified as the E1 compression mode, 
whereas the lower component may be the new mode predicted 
by the RRPA analysis of Ref.~\cite{VWR.00}. In the present 
work we suggest that the observed low-lying E1 strength may
correspond to the toroidal giant dipole resonance (TGDR).

The role of toroidal multipole form factors and moments 
in the physics of electromagnetic and weak interactions
has been extensively discussed in Refs.~\cite{DC.75}
and~\cite{DT.83}. They appear in multipole expansions
for systems containing convection and induction 
currents. In particular, the multipole expansion 
of a four-current distribution gives rise to three families 
of multipole moments: charge moments, magnetic moments and 
electric transverse moments. The later are related to 
the toroidal multipole moments and result from the expansion
of the transverse electric part of the current. The toroidal
dipole moment, in particular, describes a system of 
poloidal currents on a torus. Since the charge density 
is zero for this configuration, and all the turns of the 
torus have magnetic moments lying in the symmetry plane, 
both the charge and magnetic dipole moments of this 
configuration are equal to zero. The simplest model
is an ordinary solenoid bent into a torus.

Vortex waves in nuclei were analyzed in a hydrodynamic
model~\cite{Sem.81}. By relaxing the assumption
of irrotational motion, in this pioneering study solenoidal 
toroidal vibrations were predicted, which correspond
to the toroidal giant dipole resonance at excitation 
energy $E_x \approx (50 - 70)/A^{1/3}$ MeV.
It was suggested that the vortex excitation modes
should appear in electron backscattering.
In the framework of the time-dependent Hartree-Fock theory,
the isoscalar $1^-$ toroidal dipole states 
were also studied by analyzing the dynamics
of the moments of the Wigner transform of the density
matrix~\cite{BM.88}.

In this Letter the toroidal dipole strength distributions
are calculated in the relativistic random phase approximation 
(RRPA). The RRPA represents the small amplitude limit of the
time-dependent relativistic mean-field theory~\cite{VBR.95}.
A self-consistent calculations ensures that the same correlations which
define the ground-state properties, also determine
the behavior of small deviations from the equilibrium.
The same effective Lagrangian generates the Dirac-Hartree
single-particle spectrum and the residual particle-hole
interaction. In a number of recent applications
\cite{VWR.00,Vre.01,MGW.01,Vre.01b,RMG.01,Mawa.01}
it has been shown that, by using effective Lagrangians which in the
mean-field approximation provide an accurate description of
ground-state properties, excellent agreement with experimental
data is also found for the excitation energies of low-lying 
collective states and of giant resonances. Two points are
essential for the successful application of the RRPA in the
description of dynamical properties of finite nuclei: (i) the use
of effective Lagrangians with non-linear terms in the meson
sector, and (ii) the fully consistent treatment of the Dirac sea
of negative energy states. In particular, in Ref.~\cite{RMG.01}
it has been shown that configurations which include 
negative-energy states have an especially pronounced effect on 
isoscalar excitation modes.

In Fig.~\ref{figA}(a) we display the RRPA toroidal dipole
strength distribution for $^{208}$Pb:
\begin{equation}
R(E) = \sum_i~B^{T=0}(E1, 1_i^- \rightarrow 0_f)~
{{\Gamma^2}\over {4(E - E_i)^2 + \Gamma^2}},
\end{equation}
where $\Gamma$ is the width of the Lorentzian distribution, and
\begin{equation}
B^{T=0}(E1, 1_i^- \rightarrow 0_f)  =  \frac{1}{3} \, |
\langle 0_f || \hat{T}^{T=0}_1 || 1_i^- \rangle |^2.
\end{equation}
For the  strength distributions in Fig.~\ref{figA}
the width is $\Gamma = 1.0$ MeV.
The isoscalar toroidal dipole operator is defined~\cite{DC.75}
\begin{equation}
\hat{T}^{T=0}_{1 \mu}  = -\sqrt{\pi} \int \left [~
r^2\left ( \overrightarrow{Y}^*_{10\mu} + 
{{\sqrt{2}}\over 5} \overrightarrow{Y}^*_{12\mu}
\right )~
-  < r^2 >_{_0} \overrightarrow{Y}^*_{10\mu}\right ]\cdot \vec J(\vec r)~d^3r.
\label{TDR1}
\end{equation}
In the relativistic framework the expression for the 
the isoscalar baryon current reads
\begin{equation}
J^\nu~=~\sum_{i=1}^A \bar\psi_i\gamma^\nu\psi_i,
\label{current}
\end{equation}
where the summation is over all occupied states in the Fermi sea.
The resulting toroidal dipole operator is
\begin{equation}
\hat{T}^{T=0}_{1 \mu}  = -\sqrt{\pi} \sum_{i=1}^A \left [~
r^2_i \left ( \overrightarrow{Y}^*_{10\mu} (\Omega_i) + {{\sqrt{2}}\over 5} 
\overrightarrow{Y}^*_{12\mu} (\Omega_i) \right ) \cdot \vec \alpha_i~
-  < r^2 >_{_0} \overrightarrow{Y}^*_{10\mu} (\Omega_i) \cdot \vec \alpha_i~ 
\right ],
\label{TDR2}
\end{equation}
where $\overrightarrow{Y}_{ll^{\prime} \mu}$ denotes a vector 
spherical harmonic, and $ \vec \alpha$ are the Dirac $\alpha$-matrices.
The calculations have been performed with the
self-consistent Dirac-Hartree plus relativistic RPA. The effective
mean-field Lagrangian contains nonlinear meson self-interaction terms,
and the configuration space includes both particle-hole pairs, and pairs
formed from hole states and negative-energy states. The inclusion of
the term $- < r^2 >_{_0} \overrightarrow{Y}^*_{10\mu}$ in the
operator ensures that
the TGDR strength distributions do not contain spurious
components that correspond to the center-of-mass motion ~\cite{Sem.81}.
In Fig.~\ref{figA}(a) we compare the toroidal strength 
distributions calculated without (dashed) and with (solid) the inclusion
of this term in the operator. 

The projection of spurious center-of-mass motion components can be 
also performed by subtracting the spurious transition current, 
following a procedure similar to that adopted in Refs.~\cite{Col.00,HSZ.98}
for the IS GDR. The toroidal dipole strength distribution can be written
as
\begin{equation}
R(E) = \left| \int d^3r~\delta \vec j(\vec r)\cdot \vec T(\vec r)\right|^2\,
\end{equation}
where $\vec T(\vec r)$ is the vector toroidal operator, and 
$\delta \vec j(\vec r)$ is the transition current
\begin{equation}
\delta \vec j(\vec r) = j_-(r)\overrightarrow{Y}^*_{10\mu} (\Omega) + 
j_+(r)\overrightarrow{Y}^*_{12\mu} (\Omega).
\label{transcurr}
\end{equation}
The radial functions $j_-(r)$ and $j_+(r)$ are defined in Ref.~\cite{Serr.83}.
We have verified that identical strength distributions (solid curve in 
Fig.~\ref{figA}) are obtained when: (a) the transition current (\ref{transcurr})
is used and the toroidal operator is 
corrected by including the term $- < r^2 >_{_0} \overrightarrow{Y}^*_{10\mu}$, 
or (b) this term is not included in the operator and the spurious component
is subtracted from the transition current at each energy 
\begin{equation}
\delta \vec j(\vec r)~-~a\rho_0 \vec e_z.
\end{equation}
The second term in this expression is the spurious transition current~\cite{Ber.83}.
$\rho_0$ is the ground-state density and $\vec e_z$ denotes the unit vector in 
the direction of the center-of-mass motion. The energy dependent coefficient $a$
is determined by the condition that the integral of the transition current over 
the nucleus should vanish at each energy
\begin{equation}
\int d^3r \left(\delta \vec j(\vec r)~-~a\rho_0 \vec e_z\right)~=~0.
\end{equation}

The strength distributions in Fig.~\ref{figA} have been calculated with the
NL3~\cite{LKR.97} effective interaction. In Ref.~\cite{Vre.97}
it has been shown that isoscalar giant monopole resonances calculated
with this effective force ($K_{\rm nm} = 271.8$ MeV) are in 
excellent agreement with experimental data, and in 
Ref.~\cite{VWR.00} this interaction was used in the RRPA 
analysis of the IS GDR. 
By using effective interactions with different values of 
the nuclear matter compressibility modulus, it was shown that
only the high-energy (above 20 MeV) portion of the 
isoscalar dipole strength distribution corresponds to a compression mode.
The same effect is observed for the toroidal strength function: 
the position of the peaks in the low-energy region (below 20 MeV) depends
only weakly on the incompressibility, while the structure in the 
high-energy region is strongly affected by the choice of the 
compression modulus of the interaction. In Fig.~\ref{figA}(b) we plot
the strength function of the isoscalar dipole compression
operator~\cite{VWR.00}
\begin{equation}
\hat{Q}^{T=0}_{1 \mu}  =   \sum^{A}_{i=1} \,
\gamma_0 \, (r^3 - \frac{5}{3} \, < r^2 >_{_0}
r) \ Y_{1 \mu}(\theta_i, \varphi_i).
\label{operator}
\end{equation} 
We note that both dipole strength distributions, toroidal in Fig.~\ref{figA}(a)
and compressional in Fig.~\ref{figA}(b), display two broad structures:
one at low energies between 8 and 15 MeV, and one in the high-energy
region $25 - 30$ MeV. Obviously, one could 
expect a strong coupling between the two isoscalar $1^-$ modes.
This coupling becomes even more
evident if one rewrites the expression in square brackets of
the toroidal operator (\ref{TDR1}) as~\cite{Sem.81}
\begin{equation}
\nabla \times ( \vec r \times \nabla ) (r^3 -
\frac{5}{3} \, < r^2 >_{_0}  r) \ Y_{1 \mu},
\end{equation}
and compares it with the isoscalar dipole operator of the compression
mode (\ref{operator}). The relative position of the two resonance 
structures will, therefore, depend on the interaction between the 
toroidal and compression modes.

In Fig.~\ref{figB} we display the RRPA toroidal dipole
strength distributions in $^{90}$Zr, $^{116}$Sn, and $^{208}$Pb,
calculated with the NL3 effective interaction. In all three 
nuclei a broad, strongly fragmented structure is found in the
low-energy region, where 
"the low-lying component of the IS GDR" has been
observed~\cite{CLY.01}. The toroidal strength distributions in
this region should be compared with the experimental 
centroid energies of the "low-lying component"~\cite{CLY.01}:
$16.2\pm 0.8$ MeV for $^{90}$Zr, 
$14.7\pm 0.5$ MeV for $^{116}$Sn, and 
$12.2\pm 0.6$ MeV for $^{208}$Pb. The calculated peaks
in the high-energy region, on the other hand, correspond
to the compression mode. 
The dynamics of the solenoidal toroidal vibrations is illustrated
in Fig.~\ref{figC}, where we plot the velocity fields
for the four most pronounced
peaks of the toroidal dipole strength distributions in 
$^{116}$Sn (see Fig.~\ref{figB}).
The velocity distributions are derived from the corresponding
transition currents, following the procedure described in
Ref.~\cite{Serr.83}. A vector of
unit length is assigned to the largest velocity. All the other
velocity vectors are normalized accordingly. Since the collective flow
is axially symmetric, we plot the velocity field
in cylindrical coordinates. The $z$-axis corresponds to the
symmetry axis of a torus. We note that the two lowest peaks at
8.82 MeV  and 10.47 Mev are completely dominated
by vortex collective motion.
The velocity fields in the 
$(z,r_{\perp})$ plane correspond to poloidal currents on a
torus with vanishing inner radius. The poloidal currents determine
the dynamical toroidal moment. The high-energy peak at 30.97 MeV
displays the dynamics of dipole compression mode. 
The "squeezing" compression mode
is identified by the flow lines which
concentrate in the two "poles" on the symmetry axis.
The velocity field corresponds
to a density distribution which is being compressed in the
lower half plane, and expands in the
upper half plane. The centers of compression and expansion
are located on the symmetry axis, at approximately half
the distance between the center and the surface of the nucleus.
Finally, the intermediate peak at 17.11 MeV clearly displays the 
coupling between the toroidal and compression dipole modes.
A very similar behavior of the velocity distributions as 
function of excitation energy is also observed for $^{90}$Zr
and $^{208}$Pb.

We suggest, therefore,  that the recently observed 
"low-lying component of the isoscalar dipole mode"~\cite{CLY.01} 
might in fact correspond to the toroidal giant dipole 
resonance. By employing the fully consistent relativistic
random phase approximation, in Ref.~\cite{VWR.00} and 
in the present analysis we have shown that the 
toroidal dipole strength is concentrated
in the low-energy region around 10 MeV, while the 
isoscalar dipole excitations in the high-energy region 
above 20 MeV correspond to the "squeezing" compression mode.
States in the intermediate region display a strong mixing 
between the two dipole resonances.
The pronounced coupling between the toroidal resonance and 
the IS GDR, predicted by the RRPA calculations, might also 
explain the strong discrepancy between theory and
the experimental position of the ISGDR centroid energies
\cite{CLY.01,HSZ.98,Col.00,VWR.00}. Namely, the interaction 
causes a repulsion of the two $1^-$ isoscalar modes, i.e. it pushes
to higher energies the IS GDR and pulls to lower
energies the TGDR. This effect would explain the 
observation of Ref.~\cite{CLY.01} that: 
"The centroids of the higher (compression) mode calculated with
interactions which reproduce GMR energies are about 4 MeV higher
than the experimental centroids while the calculated centroids for
the lower mode lie 1-2 MeV below the experimental values."
\bigskip

{\bf ACKNOWLEDGMENTS}

This work has been supported in part by the
Bundesministerium f\"{u}r Bildung und Forschung under the project 06 TM 979.
D.V. and T.N. would like to acknowledge the support from the Alexander von 
Humboldt - Stiftung.

\newpage
\begin{figure}
\caption{(a)Toroidal dipole strength distributions in 
$^{208}$Pb, calculated without (dashed) and with (solid)
projection of spurious center-of-mass components.
(b) IS GDR strength distributions in $^{208}$Pb.} 
\label{figA}
\end{figure}

\begin{figure}
\caption{Toroidal dipole strength distributions in
$^{90}$Zr, $^{116}$Sn and
$^{208}$Pb, calculated with the NL3 effective interaction.}
\label{figB}
\end{figure}

\begin{figure}
\caption{Velocity distributions for the most pronounced dipole
peaks in $^{116}$Sn (see Fig. \protect\ref{figB}).
The velocity fields correspond to the peaks at 
8.82 MeV (a), 10.47 MeV (b), 17.11 MeV (c), and 30.97 MeV (d).}
\label{figC}
\end{figure}
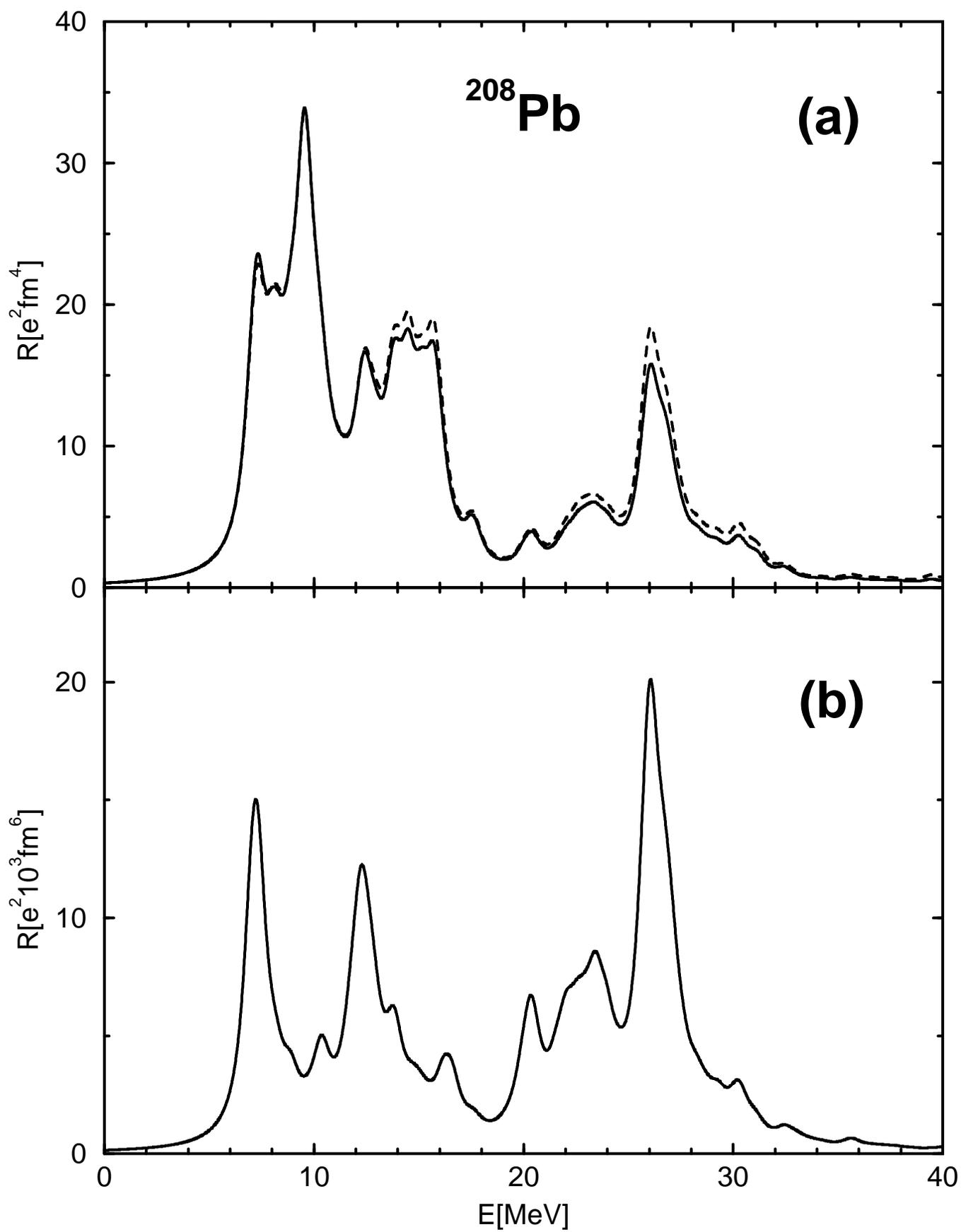
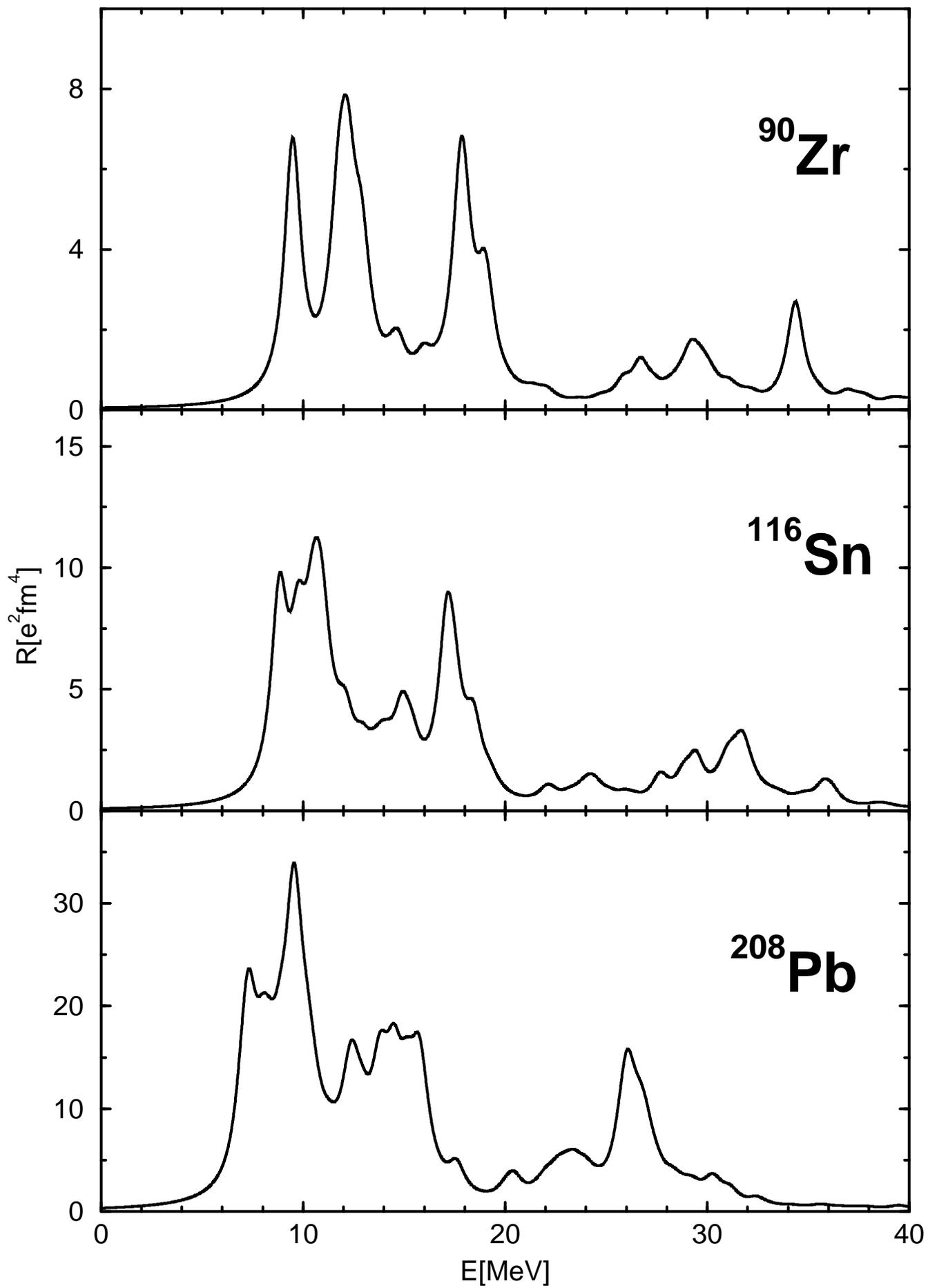
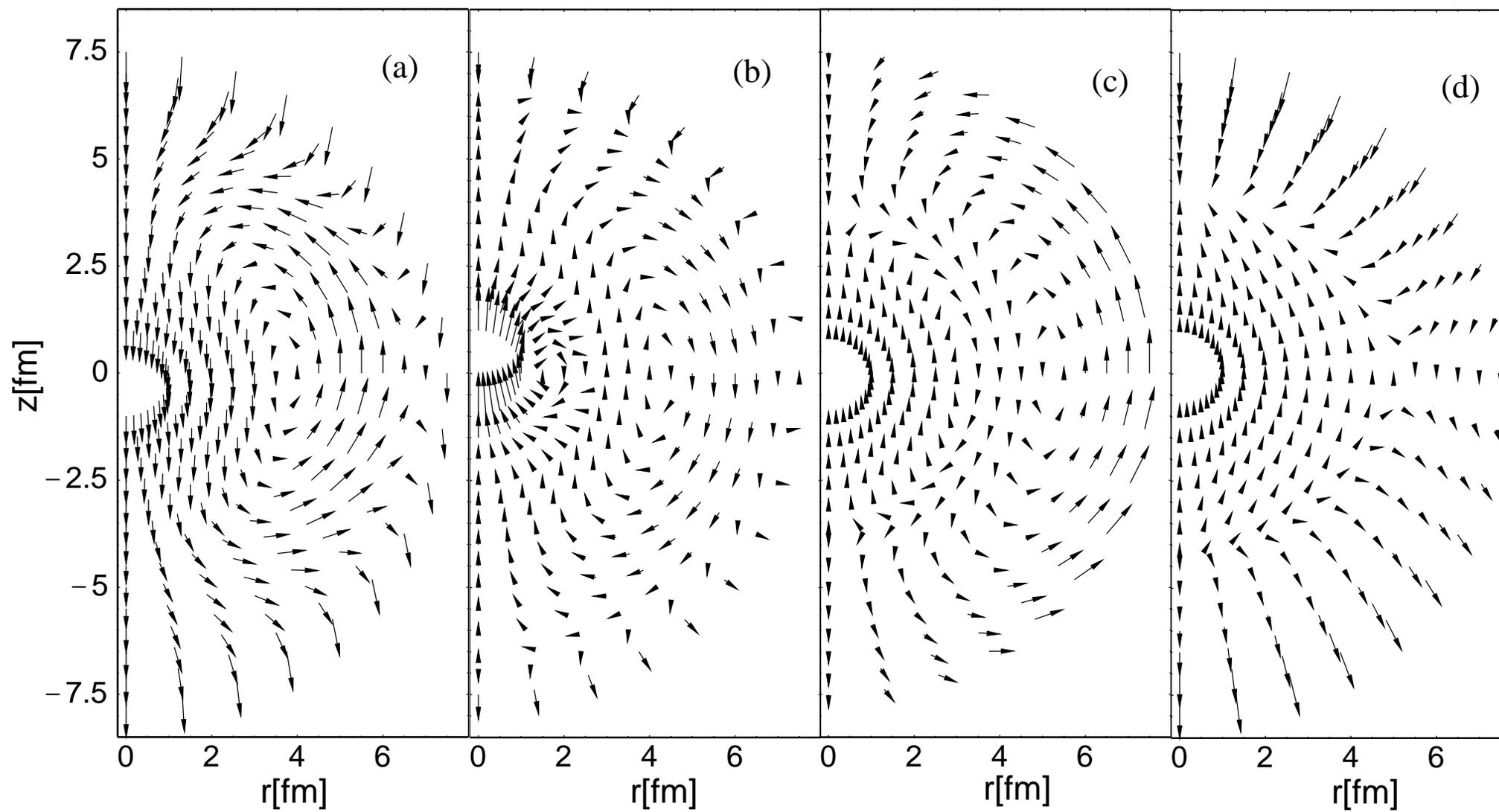
\end{document}